\documentclass[aps,twocolumn,superscriptaddress,showpacs,preprintnumbers,amsmath,amssymb]{revtex4}
\usepackage{graphicx,color,dcolumn,booktabs,bm}
\usepackage{longtable,lscape}
\usepackage{amsmath}
\usepackage{indentfirst}
\usepackage{epsfig}
\usepackage{feynmf}   
\usepackage{epstopdf}   
\usepackage{slashed}  
\usepackage{cases}
\usepackage{color}
\usepackage{multirow}
\usepackage{ulem}
\usepackage{graphicx,color,dcolumn,booktabs,bm}
\usepackage{verbatim}
\usepackage[colorlinks,linkcolor=red,anchorcolor=green,citecolor=blue]{hyperref}
\usepackage{mathrsfs}
\usepackage{rotating}
\usepackage{threeparttable}

\newcommand{\BR}{{\cal B}}

\newcommand{\pip}{\pi^+}

\newcommand{\EE}{e^+e^-}

\newcommand{\beq}{\begin{equation}}
\newcommand{\eeq}{\end{equation}}
\newcommand{\bitm}{\begin{itemize}}
\newcommand{\eitm}{\end{itemize}}

\begin{document}
\hyphenpenalty=10000


\title{\quad\\[0.1cm]\boldmath Measurements of the Branching Fractions $\BR(B^{-} \to \bar{\Lambda}_{c}^{-} \Xi_{c}^{'0})$,  $\BR(B^{-} \to \bar{\Lambda}_{c}^{-} \Xi_{c}(2645)^{0})$ and $\BR(B^{-} \to \bar{\Lambda}_{c}^{-} \Xi_{c}(2790)^{0}) $}

\noaffiliation
\affiliation{University of the Basque Country UPV/EHU, 48080 Bilbao}
\affiliation{Beihang University, Beijing 100191}
\affiliation{Brookhaven National Laboratory, Upton, New York 11973}
\affiliation{Budker Institute of Nuclear Physics SB RAS, Novosibirsk 630090}
\affiliation{Faculty of Mathematics and Physics, Charles University, 121 16 Prague}
\affiliation{Chonnam National University, Gwangju 61186}
\affiliation{University of Cincinnati, Cincinnati, Ohio 45221}
\affiliation{Deutsches Elektronen--Synchrotron, 22607 Hamburg}
\affiliation{University of Florida, Gainesville, Florida 32611}
\affiliation{Key Laboratory of Nuclear Physics and Ion-beam Application (MOE) and Institute of Modern Physics, Fudan University, Shanghai 200443}
\affiliation{Justus-Liebig-Universit\"at Gie\ss{}en, 35392 Gie\ss{}en}
\affiliation{Gifu University, Gifu 501-1193}
\affiliation{SOKENDAI (The Graduate University for Advanced Studies), Hayama 240-0193}
\affiliation{Gyeongsang National University, Jinju 52828}
\affiliation{Department of Physics and Institute of Natural Sciences, Hanyang University, Seoul 04763}
\affiliation{University of Hawaii, Honolulu, Hawaii 96822}
\affiliation{High Energy Accelerator Research Organization (KEK), Tsukuba 305-0801}
\affiliation{J-PARC Branch, KEK Theory Center, High Energy Accelerator Research Organization (KEK), Tsukuba 305-0801}
\affiliation{Forschungszentrum J\"{u}lich, 52425 J\"{u}lich}
\affiliation{IKERBASQUE, Basque Foundation for Science, 48013 Bilbao}
\affiliation{Indian Institute of Science Education and Research Mohali, SAS Nagar, 140306}
\affiliation{Indian Institute of Technology Guwahati, Assam 781039}
\affiliation{Indian Institute of Technology Hyderabad, Telangana 502285}
\affiliation{Indian Institute of Technology Madras, Chennai 600036}
\affiliation{Indiana University, Bloomington, Indiana 47408}
\affiliation{Institute of High Energy Physics, Chinese Academy of Sciences, Beijing 100049}
\affiliation{Institute of High Energy Physics, Vienna 1050}
\affiliation{Institute for High Energy Physics, Protvino 142281}
\affiliation{INFN - Sezione di Napoli, 80126 Napoli}
\affiliation{INFN - Sezione di Torino, 10125 Torino}
\affiliation{Advanced Science Research Center, Japan Atomic Energy Agency, Naka 319-1195}
\affiliation{J. Stefan Institute, 1000 Ljubljana}
\affiliation{Institut f\"ur Experimentelle Teilchenphysik, Karlsruher Institut f\"ur Technologie, 76131 Karlsruhe}
\affiliation{Kennesaw State University, Kennesaw, Georgia 30144}
\affiliation{King Abdulaziz City for Science and Technology, Riyadh 11442}
\affiliation{Korea Institute of Science and Technology Information, Daejeon 34141}
\affiliation{Korea University, Seoul 02841}
\affiliation{Kyoto University, Kyoto 606-8502}
\affiliation{Kyungpook National University, Daegu 41566}
\affiliation{LAL, Univ. Paris-Sud, CNRS/IN2P3, Universit\'{e} Paris-Saclay, Orsay 91898}
\affiliation{\'Ecole Polytechnique F\'ed\'erale de Lausanne (EPFL), Lausanne 1015}
\affiliation{P.N. Lebedev Physical Institute of the Russian Academy of Sciences, Moscow 119991}
\affiliation{Liaoning Normal University, Dalian 116029}
\affiliation{Faculty of Mathematics and Physics, University of Ljubljana, 1000 Ljubljana}
\affiliation{Ludwig Maximilians University, 80539 Munich}
\affiliation{Luther College, Decorah, Iowa 52101}
\affiliation{Malaviya National Institute of Technology Jaipur, Jaipur 302017}
\affiliation{University of Maribor, 2000 Maribor}
\affiliation{Max-Planck-Institut f\"ur Physik, 80805 M\"unchen}
\affiliation{School of Physics, University of Melbourne, Victoria 3010}
\affiliation{University of Mississippi, University, Mississippi 38677}
\affiliation{University of Miyazaki, Miyazaki 889-2192}
\affiliation{Moscow Physical Engineering Institute, Moscow 115409}
\affiliation{Moscow Institute of Physics and Technology, Moscow Region 141700}
\affiliation{Graduate School of Science, Nagoya University, Nagoya 464-8602}
\affiliation{Kobayashi-Maskawa Institute, Nagoya University, Nagoya 464-8602}
\affiliation{Universit\`{a} di Napoli Federico II, 80055 Napoli}
\affiliation{Nara Women's University, Nara 630-8506}
\affiliation{National Central University, Chung-li 32054}
\affiliation{National United University, Miao Li 36003}
\affiliation{Department of Physics, National Taiwan University, Taipei 10617}
\affiliation{H. Niewodniczanski Institute of Nuclear Physics, Krakow 31-342}
\affiliation{Nippon Dental University, Niigata 951-8580}
\affiliation{Niigata University, Niigata 950-2181}
\affiliation{University of Nova Gorica, 5000 Nova Gorica}
\affiliation{Novosibirsk State University, Novosibirsk 630090}
\affiliation{Osaka City University, Osaka 558-8585}
\affiliation{Pacific Northwest National Laboratory, Richland, Washington 99352}
\affiliation{Panjab University, Chandigarh 160014}
\affiliation{Peking University, Beijing 100871}
\affiliation{University of Pittsburgh, Pittsburgh, Pennsylvania 15260}
\affiliation{Theoretical Research Division, Nishina Center, RIKEN, Saitama 351-0198}
\affiliation{University of Science and Technology of China, Hefei 230026}
\affiliation{Showa Pharmaceutical University, Tokyo 194-8543}
\affiliation{Soongsil University, Seoul 06978}
\affiliation{University of South Carolina, Columbia, South Carolina 29208}
\affiliation{Sungkyunkwan University, Suwon 16419}
\affiliation{School of Physics, University of Sydney, New South Wales 2006}
\affiliation{Department of Physics, Faculty of Science, University of Tabuk, Tabuk 71451}
\affiliation{Tata Institute of Fundamental Research, Mumbai 400005}
\affiliation{Department of Physics, Technische Universit\"at M\"unchen, 85748 Garching}
\affiliation{Toho University, Funabashi 274-8510}
\affiliation{Earthquake Research Institute, University of Tokyo, Tokyo 113-0032}
\affiliation{Department of Physics, University of Tokyo, Tokyo 113-0033}
\affiliation{Tokyo Institute of Technology, Tokyo 152-8550}
\affiliation{Tokyo Metropolitan University, Tokyo 192-0397}
\affiliation{Virginia Polytechnic Institute and State University, Blacksburg, Virginia 24061}
\affiliation{Wayne State University, Detroit, Michigan 48202}
\affiliation{Yamagata University, Yamagata 990-8560}
\affiliation{Yonsei University, Seoul 03722}
\author{Y.~Li}\affiliation{Beihang University, Beijing 100191} 
\author{Y.~B.~Li}\affiliation{Peking University, Beijing 100871} 
\author{C.~P.~Shen}\affiliation{Key Laboratory of Nuclear Physics and Ion-beam Application (MOE) and Institute of Modern Physics, Fudan University, Shanghai 200443} 
\author{I.~Adachi}\affiliation{High Energy Accelerator Research Organization (KEK), Tsukuba 305-0801}\affiliation{SOKENDAI (The Graduate University for Advanced Studies), Hayama 240-0193} 
\author{H.~Aihara}\affiliation{Department of Physics, University of Tokyo, Tokyo 113-0033} 
\author{D.~M.~Asner}\affiliation{Brookhaven National Laboratory, Upton, New York 11973} 
\author{H.~Atmacan}\affiliation{University of South Carolina, Columbia, South Carolina 29208} 
\author{V.~Aulchenko}\affiliation{Budker Institute of Nuclear Physics SB RAS, Novosibirsk 630090}\affiliation{Novosibirsk State University, Novosibirsk 630090} 
\author{R.~Ayad}\affiliation{Department of Physics, Faculty of Science, University of Tabuk, Tabuk 71451} 
\author{V.~Babu}\affiliation{Deutsches Elektronen--Synchrotron, 22607 Hamburg} 
% ...
\collaboration{The Belle Collaboration}


\begin{abstract}
Using the data sample of 711 fb$^{-1}$ of $\Upsilon(4S)$ on-resonance data taken by the Belle detector at the KEKB asymmetric-energy electron-positron collider,
we present the first measurements of branching fractions of the decays $B^{-} \to \bar{\Lambda}_{c}^{-} \Xi_{c}^{'0}$,  $B^{-} \to \bar{\Lambda}_{c}^{-} \Xi_{c}(2645)^{0}$, and $B^{-} \to \bar{\Lambda}_{c}^{-} \Xi_{c}(2790)^{0}$.
The signal yields for these decays are extracted from the recoil
mass spectrum of the system recoiling against $\bar{\Lambda}_{c}^{-}$ baryons in selected $B^-$ candidates.
The branching fraction of $B^{-} \to \bar{\Lambda}_{c}^{-} \Xi_{c}(2790)^{0}$ is measured to be $ (1.1 \pm 0.4 \pm 0.2)\times 10^{-3}$, where the first uncertainty is statistical and the second systematic. The 90\% credibility level upper limits on $\BR(B^{-} \to \bar{\Lambda}_{c}^{-} \Xi_{c}^{'0})$ and $\BR(B^{-} \to \bar{\Lambda}_{c}^{-} \Xi_{c}(2645)^{0})$ are determined to be $6.5\times 10^{-4}$ and $7.9\times 10^{-4}$, respectively.
\end{abstract}


\maketitle
\section{\boldmath Introduction}
Charm physics is of high interest mainly due to the fact that the charm system provides a unique laboratory to study the subtle interplay of strong and weak interactions. Baryons with one charm quark and two light quarks are called charmed baryons. In the Heavy Quark Symmetry (HQS) approach~\cite{HSQ}, the two light quarks are regarded as a light diquark.
 As chiral symmetry and HQS can provide some qualitative insights into their dynamics, the
study of charmed baryons  plays an important role in improving our understanding of the quark confinement mechanism.
 The $\Xi_{c}$ charmed baryon states contain one charm quark, one strange quark, and one up or down quark.
 The ground state $\Xi_{c}^{0}$ and $\Xi_{c}^{+}$ baryons, which have spin-parity quantum numbers $J^P$ = $\frac{1}{2}^{+}$ and no internal orbital angular momentum,  are the only members of the group that decay weakly. A growing number of excited $\Xi_{c}$ states have been observed in different experiments~\cite{PDG}. However, much is still unknown about them.
 Many theoretical approaches have been used to study the excitation spectrum of $\Xi_{c}$ baryons and their decays. These models include quark models, heavy quark
$1/m_{Q}$ and $1/ N_{c}$ expansions, coupled channel model, and QCD sum rules~\cite{method0, method1,method2,method3,method4}. Through these QCD-inspired relativistic theories, the mass spectrum of excited $\Xi_{c}$ can be predicted. Recently, the masses and intrinsic widths of isodoublets of the excited $\Xi_{c}$ states $\Xi_{c}^{'}$, $\Xi_{c}(2645)$, $\Xi_{c}(2790)$, $\Xi_{c}(2815)$, and $\Xi_{c}(2980)$ were measured more precisely by Belle by analyzing their exclusive decays~\cite{xic'}.

 The decay $B^{-} \to \bar{\Lambda}_{c}^{-} \Xi_{c}^{0}$ proceeds via $b \to c\bar{c}s$ transition and has a relatively large branching fraction of the order of $10^{-3}$~\cite{PDG,xic}. Therefore, a $B$-meson factory provides an experimental research platform to investigate the $\Xi_{c}^{0}$ excitation spectrum exclusively through $B^{-} \to \bar{\Lambda}_{c}^{-} + anything$ decays. This makes it possible to search for missing excited $\Xi_{c}^{0}$ states. In addition, the measurement of their production rates is a good test for the theoretical calculation of $b \to c\bar{c}s$ transition processes.

 In this paper, we measure the branching fractions of $B^{-} \to \bar{\Lambda}_{c}^{-} \Xi_{c}^{*0}$ decays based on data collected by the Belle detector at the KEKB asymmetric-energy electron-positron collider. Here and throughout this paper, $\Xi_{c}^{*0}$ represents $\Xi_{c}^{'0}$, $\Xi_{c}(2645)^{0}$, and $\Xi_{c}(2790)^{0}$ unless otherwise stated. We use a full hadron-reconstruction algorithm~\cite{Full} to tag a $B^{+}$ signal, denoted $B_{\rm tag}^{+}$, and then reconstruct a $\bar{\Lambda}_{c}^{-}$ using its $\bar{p} K^{+} \pi^{-}$ and $\bar{p} K_{S}^{0}( K_{S}^{0} \to \pi^{+}\pi^{-})$ decay modes~\cite{charge-conjugate} from the remaining tracks. We search for peaks in the invariant mass spectrum of the system recoiling against the $\bar{\Lambda}_{c}^{-}$ baryons in the selected $B^{-} \to \bar{\Lambda}_{c}^{-} \Xi_{c}^{*0}$ candidates, to extract $\Xi_{c}^{*0}$ signal yields, from which we calculate the branching fractions of $B^{-} \to \bar{\Lambda}_{c}^{-} \Xi_{c}^{*0}$.

\section{\boldmath The data sample and the belle detector}
This analysis utilizes a data sample of 711 fb$^{-1}$ collected at the $\Upsilon(4S)$ on-resonance corresponding to $(772 \pm 11) \times10^{6}~B\bar{B}$ pairs. All the data were collected with the Belle detector~\cite{detector} operating at the KEKB asymmetric-energy $e^+e^-$ collider~\cite{collider}. The Belle detector is described in detail in Ref.~\cite{detector}. It is a large solid-angle magnetic spectrometer consisting of a silicon vertex detector, a 50-layer central drift chamber (CDC), an array of aerogel threshold Cherenkov counters (ACC), a barrel-like arrangement of time-of-flight scintillation counters (TOF), and an electromagnetic calorimeter comprised of CsI(TI) crystals  located inside a superconducting solenoid coil that provides a $1.5~\hbox{T}$ magnetic field. An iron flux return placed outside the coil is instrumented to detect $K^{0}_{L}$ mesons and to identify muons.

To optimize the signal selection criteria and to determine the signal reconstruction efficiency, Monte Carlo (MC) signal events are generated using {\sc EvtGen}~\cite{evtgen}, while $\Xi_{c}^{*0}$ inclusive decays are simulated using {\sc PYTHIA}~\cite{PYTHIA}. These events are processed by a detector simulation based on {\sc GEANT3}~\cite{geant}.
Inclusive MC samples
of $\Upsilon(4S)\to B \bar{B}$ ($B=B^+$ or $B^0$) and $e^+e^- \to q \bar{q}$
($q=u,~d,~s,~c$) events at $\sqrt{s}=10.58$ GeV corresponding to more than 3 times the integrated luminosity of the
data are used to check the backgrounds.

\section{\boldmath Common Event selection criteria}
To select the signal candidates, the following event selection criteria are applied.
For well-reconstructed charged tracks, except those from $K_{S}^{0} \to \pi^{+} \pi^{-}$ decays,
the impact parameters perpendicular to and along the beam direction with respect to the nominal interaction point (IP)
are required to be less than 1 cm and 4 cm, respectively, and the transverse momentum in the
laboratory frame is required to be larger than 0.1~GeV/$c$.
For the particle identification (PID) of a well-reconstructed charged track, information from different detector subsystems, including specific ionization in the CDC, time measurement in the TOF, and the response of the ACC, is combined to form a likelihood ${\mathcal L}_i$~\cite{pidcode} for particle species $i$, where $i$ =  $\pi$, $K$, or $p$. Tracks with $R_{K}=\mathcal{L}_{K}/(\mathcal{L}_K+\mathcal{L}_\pi)<0.4$ are identified as pions with an efficiency of 97\%, while 5\% of kaons are misidentified as pions; tracks with $R_{K}>0.6$ are identified as kaons with an efficiency of 95\%, while 4\% of pions are misidentified as kaons.
A track with
$\mathcal{R}^\pi_{p} =
\mathcal{L}_{p}/(\mathcal{L}_{p}+\mathcal{L}_\pi)
> 0.6$ and $\mathcal{R}^K_{p} =
\mathcal{L}_{p}/(\mathcal{L}_{p}+\mathcal{L}_K) >
0.6$ is identified as an (anti)proton with an efficiency of
about 97\%; fewer than $1\%$ of the pions and kaons are misidentified as (anti)protons.
With the exception of those from $K^0_S$ decays, all charged tracks
are required to be positively identified by the above procedure.

The $K_{S}^{0}$ candidates are first reconstructed from pairs of oppositely charged tracks, which are treated as pions, with a production vertex significantly separated from the average IP, then selected using a multivariate analysis using an artificial neural network~\cite{neural} based on two sets of input variables~\cite{input}.

Applying a full reconstruction algorithm of hadronic $B$-meson decays~\cite{Full} which uses a multivariate analysis based on the NeuroBayes
package, we reconstruct $B^{+}_{\rm tag}$ candidates. Each $B^{+}_{\rm tag}$ candidate has an output value $O_{\rm NN}$ from the multivariate
analysis ranging from 0 to 1. A candidate with larger $O_{\rm NN}$ is more likely to be a true $B$ meson. If multiple $B_{\rm tag}^{+}$ candidates are found in an event, only the candidate with largest $O_{\rm NN}$ is selected.
To improve the purity of the tagged side, we take $O_{\rm NN} > 0.001$,  $M_{\rm bc}^{\rm tag} > 5.27$~GeV/$c^{2}$, and $|\Delta E^{\rm tag }|< 0.04$~GeV as the signal region.
Here, $M_{\rm bc}^{\rm tag} \equiv
\sqrt{E_{\rm beam}^{2} - (\sum_i  \overrightarrow{p}^{\rm
tag}_{\!\! i})^2}$
 and $\Delta E^{\rm tag} \equiv \sum_i E^{\rm tag}_{i}
- E_{\rm beam}$, where $E_{\rm beam}\equiv \sqrt{s}/2$ is the beam energy and
$(E^{\rm tag}_{i}, \overrightarrow{p}^{\rm tag}_{\!\!i})$
is the four-momentum of the $B_{\rm tag}^+$ daughter $i$ in the $\EE$
center-of-mass system (c.m.s).
After reconstructing the $B^+_{\rm tag}$ candidate, the ${\bar{\Lambda}_c^-} \to \bar{p} K^{+} \pi^-$ and
${\bar{\Lambda}_c^-} \to \bar{p} K_{S}^{0}$ decays are reconstructed from the remaining tracks.
We perform a fit for the $\bar{\Lambda}_c^-$ decay vertex and require that $\chi^{2}_{\rm vertex}/{\rm n.d.f.} < 15$,
where ${\rm n.d.f.}$ is the number of degrees of freedom. The multi-combination rate of $\bar{\Lambda}_c^-$ candidates is 21\%. If there is more than one $\bar{\Lambda}_{c}^{-}$ candidate in an event, the candidate with the smallest $\chi^{2}_{\rm vertex}/{\rm n.d.f.}$ is selected. The $\bar{\Lambda}_{c}^{-}$ signal region is defined to be $|M_{\bar{\Lambda}_{c}^{-}} - m_{\bar{\Lambda}_{c}^{-}}| < 10$~MeV/$c^{2}$ corresponding to about $ 3\sigma$, where $\sigma$
denotes the standard deviation. Here, $M_{\bar{\Lambda}_{c}^{-}}$ is the reconstructed, and $ m_{\bar{\Lambda}_{c}^{-}}$ is the nominal, mass of the $\bar{\Lambda}_{c}^{-}$~\cite{PDG}.

\section{\boldmath  $\Xi_{c}^{'0}$, $\Xi_{c}(2645)^{0}$, and $\Xi_{c}(2790)^{0}$  signal  Extraction}

We extract the number of $\Xi_{c}^{'0}$, $\Xi_{c}(2645)^{0}$, and $\Xi_{c}(2790)^{0}$ baryons in decays of the type $B^{-} \to \bar{\Lambda}_{c}^{-} \Xi_{c}^{'0}$,  $B^{-} \to \bar{\Lambda}_{c}^{-} \Xi_{c}(2645)^{0}$, and  $B^{-} \to \bar{\Lambda}_{c}^{-} \Xi_{c}(2790)^{0}$ by fitting the recoiling mass spectrum ($M^{\rm rec}_{B_{\rm tag}^{+} \bar{\Lambda}_{c}^{-}}$). We choose 2.5~GeV/$c^2$ $ <M_{B_{\rm tag}^{+}\bar{\Lambda}_{c}^{-}}^{\rm rec} < $ 2.86~GeV/$c^2$ as the fit region. To improve the recoil mass resolution, we use $M_{B_{\rm tag}^{+}\bar{\Lambda}_{c}^{-}}^{\rm rec} \equiv M_{B_{\rm tag}^{+}\bar{\Lambda}_{c}^{-}}^{\rm miss} + M_{B_{\rm tag}^{+}} - m_{B} + M_{\bar{\Lambda}_{c}^{-}} -m_{\bar{\Lambda}_{c}^{-}}$, where $M_{B_{\rm tag}^{+}}$ is the reconstructed and $m_{B}$ is the nominal mass~\cite{PDG} of the $B^+$ meson and $M_{B_{\rm tag}^{+}\bar{\Lambda}_{c}^{-}}^{\rm miss}$ is the invariant mass recoiling against the $\bar{\Lambda}_{c}^{-}$ on the signal side, which is calculated using $\sqrt{(P_{\rm c.m.s} - P_{B^{+}_{\rm tag}} - P_{\bar{\Lambda}_{c}^{-}})^{2}}$ with $P_{\rm c.m.s},~P_{B_{\rm tag}^{+}}$, and $P_{\bar{\Lambda}_{c}^{-}}$ being four-momenta of the initial $\EE$ system, the reconstructed $B^+_{\rm tag}$ meson, and the reconstructed $\bar{\Lambda}_{c}^{-}$ baryon, respectively.

\begin{figure}[htbp]
	\begin{center}
		\includegraphics[width=7cm]{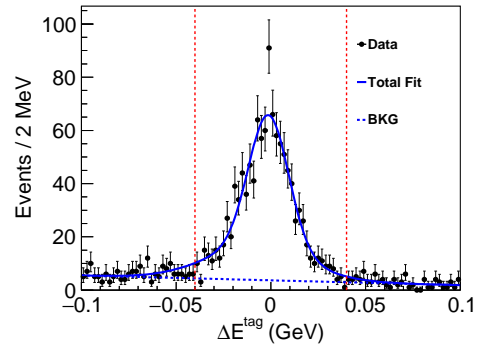} 		
	    \caption{$\Delta E^{\rm tag}$ distribution summed over the two reconstructed $\bar{\Lambda}_{c}^{-}$ decay modes. The points with error bars represent the data, the solid curve is the best-fit result, and the blue dashed curve is the fitted background. The red dashed lines show the defined
$\Delta E^{\rm tag}$ signal region.}\label{de}
	\end{center}
\end{figure}

Figure~\ref{de} shows the $\Delta E^{\rm tag}$ distribution in the $\Xi_{c}^{*0}$ signal region, i.e., 2.5~GeV/$c^2$ $ <M_{B_{\rm tag}^{+}\bar{\Lambda}_{c}^{-}}^{\rm rec} <$ 2.86~GeV/$c^2$, after applying all of the above requirements. A double-Gaussian function is used as the signal shape and the background shape is described by a first-order polynomial. Because of the small sample size, the parameters of the double-Gaussian function are fixed to the values obtained by fitting the signal MC distribution. The fit results are shown as curves in Fig.~\ref{de}. We take $|\Delta E^{\rm tag}| < 0.04$~GeV as the signal region.

\begin{figure}[htbp]
	\begin{center}
		\includegraphics[width=7.5cm]{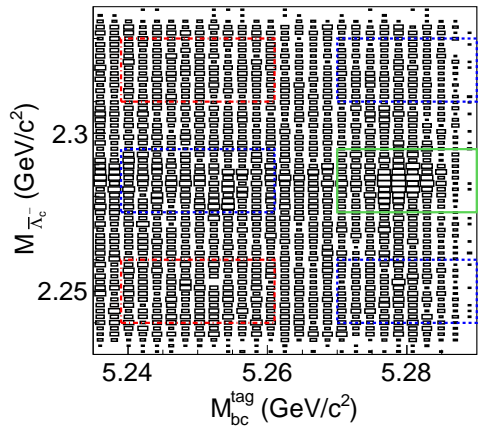}
		\caption{Scatter plot of $M_ {\bar{\Lambda}_{c}^{-}}$ of signal side versus $M_{\rm bc}^{\rm tag}$ of $B_{\rm tag}^{+}$ in the $\Xi_{c}^{*0}$ signal region, i.e., 2.5~GeV/$c^2$ $ <M_{B_{\rm tag}^{+}\bar{\Lambda}_{c}^{-}}^{\rm rec} < $ 2.86~GeV/$c^2$. The solid box shows the selected signal region, and the blue and red dashed boxes define the $M_{\rm bc}^{\rm tag}$ and $M_ {\bar{\Lambda}_{c}^{-}}$ sidebands described in the text.
		}\label{scatdata}
	\end{center}
\end{figure}

\begin{figure}[htbp]
	\includegraphics[width=9cm]{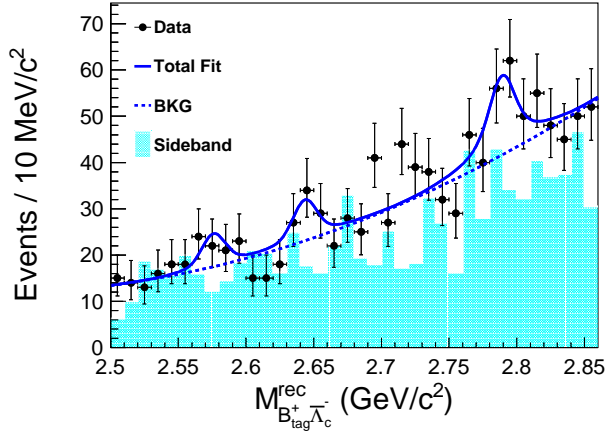}
	\caption{$M_{B_{\rm tag}^{+}\bar{\Lambda}_{c}^{-}}^{\rm rec}$ distribution of selected candidate events. The points with error bars represent the data, the solid blue curve is the best fit, the dashed curve is the fitted total background, the cyan shaded histogram is the normalized distribution of the $M_{\rm bc}^{\rm tag}$ and $M_{\bar{\Lambda}_{c}^{-}}$ sideband events (see Fig.~\ref{scatdata}).}\label{xicdata}
\end{figure}

Figure~\ref{scatdata} shows the scatter plot of $M_{\bar{\Lambda}_{c}^{-}}$ of the signal side in the $\Xi_{c}^{*0}$ signal region versus $M_{\rm bc}^{\rm tag}$ of the $B_{\rm tag}^{+}$. To check for possible peaking backgrounds, the $M_{\rm bc}^{\rm tag}$ and $M_{\bar{\Lambda}_{c}^{-}}$ sideband regions are selected as shown in Fig.~\ref{scatdata}. The normalized contribution from the $M_{\rm bc}^{\rm tag}$ and $M_{\bar{\Lambda}_{c}^{-}}$ sidebands is estimated using 50\% of the number of events in the blue dashed boxes minus 25\% of the number of events in the red dashed boxes.
Figure~\ref{xicdata} shows the $M_{B_{\rm tag}^{+}\bar{\Lambda}_{c}^{-}}^{\rm rec}$ distribution in the signal box (points with error bars) and in the sideband boxes (shaded histogram). No peaking background is found in the $M_{\rm bc}^{\rm tag}$ and $M_{\bar{\Lambda}_{c}^{-}}$ sideband
events or in the inclusive MC samples of $\Upsilon(4S) \to B \bar{B}$ and $e^+e^- \to q \bar{q}$ events.

To extract the $\Xi_{c}^{*0}$ signal yields, an unbinned maximum-likelihood fit  to the $M_{B_{\rm tag}^{+}\bar{\Lambda}_{c}^{-}}^{\rm rec}$ distribution is performed.
In this fit, the $\Xi_{c}^{'0}$ signal shape is described by a double-Gaussian function, while the $\Xi_{c}(2645)^{0}$ and $\Xi_{c}(2790)^{0}$ signal shapes are Breit-Wigner (BW) functions convolved with double-Gaussian functions.
The background is parameterized by a second-order polynomial function.
Due to the limited sample size,  the values of the parameters in double-Gaussian functions  are fixed to those obtained from the fit to the
corresponding signal MC distribution.
For $\Xi_{c}(2645)^{0}$ and $\Xi_{c}(2790)^{0}$ signal shapes,
the masses and widths of BW functions are fixed to world average values~\cite{PDG}. The fit result is shown in Fig.~\ref{xicdata}.
The difference between the fitted background level and the normalized $M_{\rm bc}^{\rm tag}$ and $M_{\bar{\Lambda}_{c}^{-}}$
sidebands is due to the contribution from other multibody $B^-$ decay modes with a $\bar{\Lambda}_{c}^{-}$,
for example, $B^- \to K^- \Lambda_{c}^{+} \bar{\Lambda}_{c}^{-}$.

The numbers of fitted $\Xi_{c}^{'0}$, $\Xi_{c}(2645)^{0}$, and $\Xi_{c}(2790)^{0}$ signal events are $N_{\Xi_{c}^{'0}} = 17.9 \pm 10.4$, $N_{\Xi_{c}(2645)^{0}} = 24.1 \pm 13.0$,  and  $N_{\Xi_{c}(2790)^{0}} = 59.9 \pm 22.5$ with statistical significances of $1.7\sigma$, $1.9\sigma$ and $3.1\sigma$, respectively. Here, the statistical significances are defined as $\sqrt{-2\ln(\mathcal{L}_{0}/\mathcal{L}_\text{max})}$, where $\mathcal{L}_{0}$ and $\mathcal{L}_\text{max}$ are the maximized likelihoods without and with a signal component, respectively\cite{CLF,CLF2}. The $\Xi_{c}(2790)^{0}$ signal significance
becomes 3.0$\sigma$ when systematic uncertainties are included (see below).

Then the branching fractions are
 \begin{align*}
\BR_{\bar{\Lambda}_{c}^{-}\Xi_{c}^{'0}}
&=\frac{N_{\Xi_{c}^{'0}}}{2 N_{B^{-}}(\varepsilon_{\bar{p}K^{+}\pi^{-}}^{\Xi_{c}^{'0}}\BR_{1}+\varepsilon_{\bar{p}K_{s}^{0}}^{\Xi_{c}^{'0}}\BR_{2})}
\\&=(3.4\pm2.0)\times10^{-4},
\end{align*}
\begin{align*}
\BR_{\bar{\Lambda}_{c}^{-}\Xi_{c}(2645)^{0}}&=\frac{N_{\Xi_{c}(2645)^{0}}}{2 N_{B^{-}}(\varepsilon_{\bar{p}K^{+}\pi^{-}}^{\Xi_{c}(2645)^{0}}\BR_{1}+\varepsilon_{\bar{p}K_{s}^{0}}^{\Xi_{c}(2645)^{0}}\BR_{2})}\\&=(4.4\pm2.4)\times 10^{-4},
\end{align*}
and
\begin{align*}
\BR_{\bar{\Lambda}_{c}^{-}\Xi_{c}(2790)^{0}}&= \frac{N_{\Xi_{c}(2790)^{0}}}{2 N_{B^{-}}(\varepsilon_{\bar{p}K^{+}\pi^{-}}^{\Xi_{c}(2790)^{0}}\BR_{1}+\varepsilon_{\bar{p}K_{s}^{0}}^{\Xi_{c}(2790)^{0}}\BR_{2})}\\&=(1.1\pm0.4)\times 10^{-3},
\end{align*}
where
$\BR_{\bar{\Lambda}_{c}^{-}\Xi_{c}^{'0}}$ =$\BR(B^{-}\to\bar{\Lambda}_{c}^{-}\Xi_{c}^{'0})$,
$\BR_{\bar{\Lambda}_{c}^{-}\Xi_{c}(2645)^{0}}$=$\BR(B^{-}\to\bar{\Lambda}_{c}^{-}\Xi_{c}(2645)^{0})$,
$\BR_{\bar{\Lambda}_{c}^{-}\Xi_{c}(2790)^{0}}$=$\BR(B^{-}\to\bar{\Lambda}_{c}^{-}\Xi_{c}(2790)^{0})$,
$N_{B^{-}}=N_{\Upsilon(4S)}\BR(\Upsilon(4S)\to B^+B^-)$ with $N_{\Upsilon(4S)}$ being the number of accumulated $\Upsilon(4S)$ events. We use a value of 0.514 for $\BR(\Upsilon(4S)\to B^+B^-)$~\cite{PDG}; $\BR_{1} = \BR(\bar{\Lambda}_{c}^{-}\to \bar{p}K^{+}\pi^{-}),~\BR_{2} = \BR(\bar{\Lambda}_{c}^{-}\to\bar{p}K_S^0 )\BR(K_S^0 \to \pi^+\pi^-)$, $\BR(\bar{\Lambda}_{c}^{-} \to \bar{p} K^{+} \pi^{-})$, $\BR(\bar{\Lambda}_{c}^{-} \to \bar{p}K_S^0)$, and $\BR(K_S^0 \to \pi^+\pi^-)$ are the branching fractions of $\bar{\Lambda}_{c}^{-} \to \bar{p} K^{+} \pi^{-}$, $\bar{\Lambda}_{c}^{-} \to \bar{p}K_S^0$, and $K_S^0 \to \pi^+\pi^-$~\cite{PDG}, respectively.
$\varepsilon_{ \bar{p}K^{+}\pi^{-}}^{\Xi_{c}^{*0}}$ and $\varepsilon_{\bar{p}K_{s}^{0}}^{\Xi_{c}^{*0}}$
are the detection efficiencies of different $\bar{\Lambda}_{c}^{-}$ decay modes which are obtained from fits to the signal MC samples and are listed in Table~\ref{eff}.

\begin{table}[htbp]
	\begin{center}
		\caption{\label{eff} The detection efficiencies $\varepsilon_{ \bar{p}K^{+}\pi^{-}}^{\Xi_{c}^{*0}}$ and $\varepsilon_{\bar{p}K_{s}^{0}}^{\Xi_{c}^{*0}}$
including $B^+_{\rm tag}$ meson for the studied $\bar{\Lambda}_{c}^{-}$ decay modes as obtained from MC simulated $B^- \to \bar{\Lambda}_{c}^{-}  \Xi_{c}^{*0}$ processes.  All the uncertainties here are statistical only.}
		\renewcommand\arraystretch{1.8}
		\begin{tabular}{cccc}	
	
			\hline
			\hline
			$\Xi_{c}^{*0}$ type  & $\Xi_{c}^{'0}$ &  $\Xi_{c}(2645)^{0}$ & $\Xi_{c}(2790)^{0}$\\
		    \hline
			$\varepsilon_{ \bar{p}K^{+}\pi^{-}}^{\Xi_{c}^{*0}}$(\%)     & $0.09 \pm 0.003$ &$0.09 \pm 0.003$  & $0.09 \pm 0.003$  \\
					
			$\varepsilon_{\bar{p}K_{s}^{0}}^{\Xi_{c}^{*0}}$(\%)          & $0.13 \pm 0.003$ & $0.14 \pm 0.004$  & $0.15 \pm 0.004$ \\
			\hline
			\hline	
					
		\end{tabular}
	\end{center}
\end{table}

Since the statistical significances of $\Xi_{c}^{'0}$ and $\Xi_{c}(2645)^{0}$ are less than 3$\sigma$, Bayesian upper limits at 90\% credibility level (C.L.)~\cite{CL} assuming a uniform prior probability for $\BR(B^-\to\bar{\Lambda}_{c}^{-}\Xi_{c}^{'0})$ and $\BR(B^-\to\bar{\Lambda}_{c}^{-}\Xi_{c}(2645)^{0})$ (denoted $\BR^{\rm up}$) are determined by solving the equation $$\int_0^{\BR^{\rm up}}\mathcal{L}(\BR)d\BR /\int_0^{1}\mathcal{L}(\BR)d\BR=0.9,$$ where 
$$\BR= N_{\Xi_{c}^{*0}}/2N_{B^{-}}(\varepsilon_{\bar{p}K^{+}\pi^{-}}^{\Xi_{c}^{*0}}\BR_{1}+\varepsilon_{\bar{p}K_{s}^{0}}^{\Xi_{c}^{*0}}\BR_{2})$$
is the assumed  branching fraction for $B^-\to\bar{\Lambda}_{c}^{-}\Xi_{c}^{'0}$ or $B^-\to\bar{\Lambda}_{c}^{-}\Xi_{c}(2645)^{0}$, and $\mathcal{L} (\BR)$ is the corresponding likelihood of the data. $N_{\Xi_{c}^{*0}}$ is the fitted signal yield of $\Xi_{c}^{'0}$ or $\Xi_{c}(2645)^{0}$.  Taking into account the systematic uncertainty discussed below, the likelihoods are convolved with a Gaussian function whose width equals the corresponding total systematic uncertainty. The 90\% C.L. upper limits with systematic uncertainties included on $\BR(B^{-} \to \bar{\Lambda}_{c}^{-} \Xi_{c}^{'0})$ and  $\BR(B^{-} \to \bar{\Lambda}_{c}^{-} \Xi_{c}(2645)^{0})$ are determined to be $6.5\times 10^{-4}$ and $7.9\times 10^{-4}$, respectively.

\begin{table*}[htbp]
	\begin{center}
		\caption{\label{tab:summary} Summary of the fitted signal yields ($N_{\rm sig}$),
         branching fractions (90\% C.L. upper limits), and
         statistical signal significances ($\sigma$) for $B^{-} \to \bar{\Lambda}_{c}^{-} \Xi_{c}^{'0}$,  $B^{-} \to \bar{\Lambda}_{c}^{-} \Xi_{c}(2645)^{0}$, and  $B^{-} \to \bar{\Lambda}_{c}^{-} \Xi_{c}(2790)^{0}$.  All the uncertainties here are
         statistical only. }
		\renewcommand\arraystretch{1.3}
		\begin{tabular}{cccc}
			\hline\hline
		       &  $N_{\rm sig}$ 
			& $\BR(B^{-} \to \bar{\Lambda}_{c}^{-} \Xi_{c}^{*0})$ [Upper Limit]&  Significance ($\sigma$) \\
			\hline
			$\Xi_{c}^{'0}$       & $ 17.9 \pm 10.4 $  & $(3.4 \pm 2.0) \times 10^{-4}$ [$6.5\times 10^{-4}$] & 1.7  \\
		
			$\Xi_{c}(2645)^{0}$  & $ 24.1 \pm 13.0 $  & $(4.4 \pm 2.4) \times 10^{-4}$ [$7.9\times 10^{-4}$] & 1.9 \\
		
			$ \Xi_{c}(2790)^{0}$ & $ 59.9 \pm 22.5 $  & $(1.1 \pm 0.4) \times 10^{-3}$ & 3.1  \\
			\hline\hline
			
		\end{tabular}
	\end{center}
\end{table*}

Table~\ref{tab:summary} summarizes the fitted results, branching fractions, and statistical significances
for $B^{-} \to \bar{\Lambda}_{c}^{-} \Xi_{c}^{'0}$,  $B^{-} \to \bar{\Lambda}_{c}^{-} \Xi_{c}(2645)^{0}$, and  $B^{-} \to \bar{\Lambda}_{c}^{-} \Xi_{c}(2790)^{0}$. The uncertainties shown are statistical only.

\section{\boldmath Systematic Uncertainties}
There are several sources of systematic uncertainties for the branching fraction measurements as listed in Table~\ref{tab:err2},
including the reconstruction-efficiency-related sources, the fit uncertainty, the $\Lambda_{c}$ decay branching fractions,
the $B$ meson tag efficiency, and the total number of $B\bar{B}$ events.

The reconstruction-efficiency-related uncertainties include those for tracking efficiency (0.35\% per track),
particle identification efficiency (1.44\% per kaon, 0.86\% per pion, and range from 2.13\% to 3.13\% per proton), as well as  momentum-weighted $K_{S}^{0}$ selection efficiency (1.1\%)~\cite{kserr}. Here, the systematic uncertainty due to the $K_{S}^{0}$ selection depends on the $K_{S}^{0}$ momuntum and was determined using a control sample of $D^{*+} \to D^{0}(K_{S}^{0}\pi^{0})\pip$.
For the three branching-fraction measurements, the individual reconstruction-efficiency-related uncertainties from two different $\bar{\Lambda}_c^-$ decay channels are added linearly weighted by the product of the detection efficiency and $\bar{\Lambda}_c^-$ partial decay width. Then those uncertainties are summed in quadrature to be the final uncertainties related to the efficiency of the reconstruction, yielding 3.1 to 3.5\%, depending on the specific decay mode.

We estimate the systematic uncertainties associated with the fit
by changing the order of the background polynomial, by changing the range of the fit,
and by enlarging the mass resolution by 10\%.
The observed deviations are taken as
systematic uncertainties.
The masses of $\Xi_{c}(2790)^{0}$ and $\Xi_{c}(2815)^{0}$ are rather close,
and no $\Xi_{c}(2815)^{0}$ signal peak can be seen. The $\Xi_{c}(2815)^{0}$ signal significance
is only 0.4$\sigma$ if it is added in the fit. So, we take the difference of the number
of $\Xi_{c}(2790)^{0}$ signal events as the systematic uncertainty due to the possible contribution of $\Xi_{c}(2815)^{0}$ from
$B^{-} \to \bar{\Lambda}_{c}^{-} \Xi_{c}(2815)^{0}$.
Finally, all the above uncertainties are summed in quadrature and
the sums are taken as the systematic uncertainties associated with the fit.

Uncertainties for the $\bar{\Lambda}_{c}^{-}$ decay branching fractions are due to
$\BR(\bar{\Lambda}_{c}^{-} \to f_{i}) = \Gamma_{i} \times \BR(\bar{\Lambda}_{c}^{-} \to \bar{p} K^{+} \pi^{-})$;
here $ \Gamma_{i}= {\BR(\bar{\Lambda}_{c}^{-} \to f_{i})}/{\BR(\bar{\Lambda}_{c}^{-} \to \bar{p} K^{+} \pi^{-})}$
and $f_{i}$ denotes the different $\bar{\Lambda}_{c}^{-}$ decay modes.
Uncertainties on $\BR(\bar{\Lambda}_{c}^{-} \to \bar{p} K^{+} \pi^{-})$ and
$\Gamma(\bar{\Lambda}_{c}^{-} \to \bar{p} K_{s}^{0})/\Gamma(\bar{\Lambda}_{c}^{-} \to \bar{p} K^{+} \pi^{-})$ are
taken from Ref.~\cite{PDG}. The final uncertainties on the two $\bar{\Lambda}_{c}^{-}$ partial decay widths
are summed in quadrature with the detection efficiency as a weighting factor.
The uncertainty due to the $B$ meson tagging efficiency is 4.2\%~\cite{btag_err}.
The uncertainty on $\BR(\Upsilon(4S)\to B^+B^-)$ is 1.2\%~\cite{PDG}. The systematic uncertainty
on $N_{\Upsilon(4S)}$ is 1.37\%.
The sources of uncertainty summarized in Table~\ref{tab:err2} are assumed to be independent
and thus are added in quadrature to obtain the total systematic uncertainty.

\begin{table*}[htbp]
	\caption{\label{tab:err2} Summary of the relative systematic uncertainties on the branching fraction measurements
(\%) for $B^{-} \to \bar{\Lambda}_{c}^{-} \Xi_{c}^{'0}$,  $B^{-} \to \bar{\Lambda}_{c}^{-} \Xi_{c}(2645)^{0}$, and  $B^{-} \to \bar{\Lambda}_{c}^{-} \Xi_{c}(2790)^{0}$.}
   \begin{tabular}{cccccccc}
    \hline\hline Observable &Efficiency&Fit&$\bar{\Lambda}_{c}^{-}$  decays&$B_{\rm tag}$ &$N_{B^{\pm}}$ &Total&Measured value\\
	\hline
		$\BR(B^{-} \to \bar{\Lambda}_{c}^{-} \Xi_{c}^{'0})$      & 3.1 & 10.0   & 5.5 & 4.2 & 1.8 & 12.6 &  $(3.4 \pm 2.0 \pm 0.4)\times 10^{-4}$\\
		$\BR(B^{-} \to \bar{\Lambda}_{c}^{-} \Xi_{c}(2645)^{0})$ & 3.3 & 8.1   & 5.5 & 4.2 & 1.8 & 11.3 &  $(4.4 \pm 2.4 \pm 0.5)\times 10^{-4}$\\
		$\BR(B^{-} \to \bar{\Lambda}_{c}^{-} \Xi_{c}(2790)^{0})$ & 3.5 & 11.0  & 5.5 & 4.2 & 1.8 & 13.6 &  $(1.1 \pm 0.4 \pm 0.2)\times 10^{-3}$\\
		\hline\hline
	\end{tabular}
\end{table*}

\section{\boldmath conclusion}
Using the 711 fb$^{-1}$ data sample taken at the $\Upsilon(4S)$ resonance that corresponds to $(772 \pm 11) \times10^{6}~B\bar{B}$ pairs accumulated with the Belle detector at the KEKB asymmetric-energy $e^+e^-$ collider, we present the  first  measurements of the branching fractions of the decays $B^{-} \to \bar{\Lambda}_{c}^{-} \Xi_{c}^{'0}$,  $B^{-} \to \bar{\Lambda}_{c}^{-} \Xi_{c}(2645)^{0}$, and  $B^{-} \to \bar{\Lambda}_{c}^{-} \Xi_{c}(2790)^{0}$ with $\Xi_{c}^{*0} \to anything$ and the $\bar{\Lambda}_{c}^{-}$ candidates reconstructed via their $\bar{p} K^+ \pi^-$ and $\bar{p} K_S^0$ decay modes. The branching fractions are measured to be $$\BR(B^{-} \to \bar{\Lambda}_{c}^{-} \Xi_{c}^{'0})=(3.4 \pm 2.0 \pm  0.4)\times 10^{-4},$$ $$\BR(B^{-} \to \bar{\Lambda}_{c}^{-} \Xi_{c}(2645)^{0})=(4.4 \pm 2.4 \pm 0.5)\times 10^{-4},$$ and $$\BR(B^{-} \to \bar{\Lambda}_{c}^{-} \Xi_{c}(2790)^{0})=(1.1 \pm 0.4 \pm  0.2)\times 10^{-3},$$ with statistical significances of $ 1.7\sigma$, $ 1.9\sigma$, and $ 3.1\sigma$, respectively. Since the statistical significances are less than 3$\sigma$
for $B^{-} \to \bar{\Lambda}_{c}^{-} \Xi_{c}^{'0}$ and $B^{-} \to \bar{\Lambda}_{c}^{-} \Xi_{c}(2645)^{0}$, the 90\% C.L. upper limits on $\BR(B^{-} \to \bar{\Lambda}_{c}^{-} \Xi_{c}^{'0})$ and  $\BR(B^{-} \to \bar{\Lambda}_{c}^{-} \Xi_{c}(2645)^{0})$  are determined to be $6.5\times 10^{-4}$ and  $7.9\times 10^{-4}$, respectively, with systematic uncertainties included.

We thank the KEKB group for the excellent operation of the
accelerator; the KEK cryogenics group for the efficient
operation of the solenoid; and the KEK computer group, and the Pacific Northwest National
Laboratory (PNNL) Environmental Molecular Sciences Laboratory (EMSL)
computing group for strong computing support; and the National
Institute of Informatics, and Science Information NETwork 5 (SINET5) for
valuable network support.  We acknowledge support from
the Ministry of Education, Culture, Sports, Science, and
Technology (MEXT) of Japan, the Japan Society for the
Promotion of Science (JSPS), and the Tau-Lepton Physics
Research Center of Nagoya University;
the Australian Research Council including grants
DP180102629, 
DP170102389, 
DP170102204, 
DP150103061, 
FT130100303; 
Austrian Science Fund (FWF);
the National Natural Science Foundation of China under Contracts
No.~11435013,  
No.~11475187,  
No.~11521505,  
No.~11575017,  
No.~11675166,  
No.~11705209,  
No. 11761141009, 
No. 11975076;
Key Research Program of Frontier Sciences, Chinese Academy of Sciences (CAS), Grant No.~QYZDJ-SSW-SLH011; 
the  CAS Center for Excellence in Particle Physics (CCEPP); 
the Shanghai Pujiang Program under Grant No.~18PJ1401000;  
the Ministry of Education, Youth and Sports of the Czech
Republic under Contract No.~LTT17020;
the Carl Zeiss Foundation, the Deutsche Forschungsgemeinschaft, the
Excellence Cluster Universe, and the VolkswagenStiftung;
the Department of Science and Technology of India;
the Istituto Nazionale di Fisica Nucleare of Italy;
National Research Foundation (NRF) of Korea Grant
Nos.~2016R1\-D1A1B\-01010135, 2016R1\-D1A1B\-02012900, 2018R1\-A2B\-3003643,
2018R1\-A6A1A\-06024970, 2018R1\-D1A1B\-07047294, 2019K1\-A3A7A\-09033840,
2019R1\-I1A3A\-01058933;
Radiation Science Research Institute, Foreign Large-size Research Facility Application Supporting project, the Global Science Experimental Data Hub Center of the Korea Institute of Science and Technology Information and KREONET/GLORIAD;
the Polish Ministry of Science and Higher Education and
the National Science Center;
the Ministry of Science and Higher Education of the Russian Federation, Agreement 14.W03.31.0026; 
the Slovenian Research Agency;
Ikerbasque, Basque Foundation for Science, Spain;
the Swiss National Science Foundation;
the Ministry of Education and the Ministry of Science and Technology of Taiwan;
and the United States Department of Energy and the National Science Foundation.


\begin{thebibliography}{**}
\bibitem{HSQ} N.~Isgur and M.B.~Wise, Phys. Rev. Lett. {\bf 66}, 1130 (1991).
\bibitem{PDG} M.~Tanabashi {\it et al.} (Particle Data Group), Phys. Rev. D {\bf 98}, 030001 (2018).
\bibitem{method0} D.~Ebert, R.N.~Faustov, and V.O.~ Galkin, Phys. Rev. D {\bf 72}, 034026 (2005).
\bibitem{method1} C.~Semay, F.~Buisseret, and F.~Stancu, Phys. Rev. D {\bf 78}, 076003 (2008).
\bibitem{method2} O.~Romanets, L.~Tolos, C.~Garcia-Recio, J.~Nieves, L.L.~Salcedo, and R.G.E.~ Timmermans, Phys. Rev. D {\bf 85}, 114032 (2012).
\bibitem{method3} J.R.~Zhang and M.Q.~Huang, Phys. Rev. D {\bf 78}, 094015 (2008).
\bibitem{method4} H.X.~Chen, W.~Chen, Q.~Mao, A.~Hosaka, X.~Liu, and S.L.~Zhu, Phys. Rev. D {\bf 91}, 054034 (2015).
\bibitem{xic'} J.~Yelton {\it et al.} (Belle Collaboration), Phys. Rev. D {\bf 94}, 052011 (2016).
\bibitem{xic} Y.B.~Li {\it et al.} (Belle Collaboration), Phys. Rev. Lett. {\bf 122}, 082001 (2019).
\bibitem{Full} M.~Feindt, F.~Keller, M.~Kreps, T.~Kuhr, S.~Neubauer, D.~Zander, and A.~Zupanc, Nucl. Instrum. Methods Phys. Res., Sect. A {\bf 654}, 432 (2011).
\bibitem{charge-conjugate} Inclusion of charge-conjugate states is implicit unless otherwise stated.
\bibitem{detector} A.~Abashian {\it et al}. (Belle Collaboration), Nucl. Instrum.
Methods Phys. Res., Sect. A {\bf 479}, 117 (2002); also, see
detector section in J. Brodzicka {\it et al.}, Prog. Theor. Exp. Phys. {\bf 2012}, 04D001 (2012).
\bibitem{collider} S.~Kurokawa and E.~Kikutani, Nucl. Instrum. Methods
Phys. Res., Sect. A {\bf 499}, 1 (2003), and other papers included
in this volume; T. Abe {\it et al.}, Prog. Theor. Exp. Phys. {\bf 2013},
03A001 (2013), and references therein.
\bibitem{evtgen} D.J.~Lange, Nucl. Instrum. Methods Phys. Res., Sect. A {\bf 462}, 152 (2001).
\bibitem{PYTHIA} T.~Sj{\"{o}}strand {\it et al.}, Comput. Phys. Commun. {\bf 135}, 238 (2001).
\bibitem{geant} R.~Brun {\it et al.}, GEANT 3: user's guide Geant 3.10, Geant 3.11, CERN Report No. DD/EE/84-1, 1984.
\bibitem{pidcode} E.~Nakano, Nucl. Instrum. Methods Phys. Res., Sect. A {\bf 494}, 402 (2002).
\bibitem{neural} M.~Feindt and U.~Kerzel, Nucl. Instrum. Methods Phys. Res., Sect. A {\bf 559}, 190 (2006).
\bibitem{input} H.~Nakano, Ph.D Thesis, Tohoku University (2014) Chapter 4, unpublished,
https://tohoku.repo.nii.ac.jp/?action=pages\_view\_main\&
active\_action=repository\_view\_main\_item\_detail\&item\_id
=70563\&item\_no=1\&page\_id=33\&block\_id=38.  
\bibitem{CLF}  S. S. Wilks, Ann. Math. Stat. {\bf9}, 60 (1938).
\bibitem{CLF2}  G. Cowan {\it et al.}, Eur. Phys. J. C {\bf 73}, 2501 (2013).
\bibitem{CL}{In common HEP usage, this Bayesian interval has been reported as ``confidence interval", which is a frequentist-statistics term.}
\bibitem{kserr} N.~Dash {\it et al.} (Belle Collaboration), Phys. Rev. Lett. {\bf 119}, 171801 (2017).
\bibitem{btag_err} A.~Sibidanov {\it et al.} (Belle Collaboration), Phys. Rev. D {\bf 88}, 032005 (2013).
\end{thebibliography}
\end{document}